\documentclass{aa}  
\usepackage{textcomp}
\usepackage[utf8]{inputenc}
\usepackage{lscape}
\usepackage{lastpage}
\usepackage{booktabs}
\usepackage{amsmath,amssymb}
\usepackage{mathrsfs}	
\usepackage{multicol}

\usepackage{graphicx}
\usepackage{txfonts}
\usepackage{hyperref}
\hypersetup{
    colorlinks   = true, 
    urlcolor     = cyan, 
    linkcolor    = blue, 
    citecolor   = blue 
}
\usepackage{placeins}
\usepackage{xcolor}
\usepackage{caption}

\graphicspath{plots}

{\newif\ifnotend
\notendtrue
\def\veclist{ABCDEFGHIJKLMNOPQRSTUVWXYZabcdefghijklmnopqrstuvwxyz.}
\def\top#1#2.{#1}
\def\tail#1#2.{#2.}
\loop\expandafter\xdef\csname bb\expandafter\top\veclist\endcsname%
{{\noexpand\bf\expandafter\top\veclist}}
\edef\veclist{\expandafter\tail\veclist}
\if\veclist.\notendfalse\fi\ifnotend\repeat}
\def\d{{\rm d}}

\newcommand{\kpc}   {\,{\rm kpc}}
\newcommand{\pc}    {\,{\rm pc}}

\newcommand{\Msun}  {\,M_{\odot}}
\newcommand{\Gyr}   {\,{\rm Gyr}}

\newcommand{\kms}   {\,{\rm km\,s^{-1}}}
\newcommand{\asec}  {\,{\rm arcsec}}
\newcommand{\amin}  {\,{\rm arcmin}}
\newcommand{\cm}    {\,{\rm cm}}

\newcommand{\GeV}     {\,{\rm GeV}}

\newcommand{\fJ}    {f({\boldsymbol J})}
\newcommand{\fJi}   {f_i({\boldsymbol J})}

\newcommand{\hJi}   {h_i({\boldsymbol J})}

\newcommand{\gJi}   {g_i({\boldsymbol J})}

\newcommand{\fni}   {f_i}
\newcommand{\Mi}    {M_i}
\newcommand{\Mst}   {M_\star}
\newcommand{\Mdm}   {M_{\rm DM}}
\newcommand{\MBH}   {M_{\rm BH}}

\newcommand{\Ji}    {J_i}
\newcommand{\Jst}   {J_{\star}}
\newcommand{\Jdm}   {J_{\rm DM}}

\newcommand{\Jti}    {J_{{\rm t}, i}}
\newcommand{\Jtst}   {J_{{\rm t},\star}}
\newcommand{\Jtdm}   {J_{\rm t, DM}}

\newcommand{\Jci}    {J_{{\rm c}, i}}
\newcommand{\Jcst}   {J_{{\rm c},\star}}
\newcommand{\Jcdm}   {J_{\rm c, DM}}

\newcommand{\alphai}    {\alpha_{i}}
\newcommand{\alphast}   {\alpha_\star}
\newcommand{\alphadm}   {\alpha_{\rm DM}}

\newcommand{\etai}   {\eta_{i}}
\newcommand{\etast}  {\eta_\star}
\newcommand{\etadm}  {\eta_{\rm DM}}

\newcommand{\Bi}   {B_{i}}
\newcommand{\Bst}  {B_\star}
\newcommand{\Bdm}  {B_{\rm DM}}

\newcommand{\Gammai}   {\Gamma_{i}}
\newcommand{\Gammast}  {\Gamma_\star}
\newcommand{\Gammadm}  {\Gamma_{\rm DM}}

\newcommand{\hri}  {h_{r,i}}

\newcommand{\hzst} {h_{z,\star}}
\newcommand{\hzdm} {h_{z,{\rm DM}}}

\newcommand{\gri}  {g_{r,i}}

\newcommand{\gzst} {g_{z,\star}}
\newcommand{\gzdm} {g_{z,{\rm DM}}}

\newcommand{\Jr}    {J_r}
\newcommand{\Jphi}  {J_{\phi}}
\newcommand{\Jz}    {J_z}

\newcommand{\sigmar}{\sigma_r}
\newcommand{\sigmat}{\sigma_\theta}
\newcommand{\tbeta}  {\tilde{\beta}}

\newcommand{\JJ}    {\boldsymbol{J}}

\newcommand{\Phitot}{\Phi_{\rm tot}}
\newcommand{\PhiBH} {\Phi_{\rm BH}}
\newcommand{\Phist} {\Phi_\star}
\newcommand{\Phidm} {\Phi_{\rm DM}}
\newcommand{\Phii}  {\Phi_i}

\newcommand{\vlos}  {v_{\rm los}}

\newcommand{\rinfl} {r_{\rm infl}}
\newcommand{\Reff}  {R_{\rm eff}}
\newcommand{\rperi} {r_{\rm peri}}

\newcommand{\gammaCL} {\gamma_{150}}
\newcommand{\rhoCL} {\rho_{150}}
\newcommand{\rhodm} {\rho_{\rm DM}}
\newcommand{\Mdyn} {M_{\rm dyn}}

\newcommand{\rc}    {r_{\rm c}}
\newcommand{\rt}    {r_{\rm t}}
\newcommand{\MV}    {M_{\rm V}}

\newcommand{\dd}    {{\rm d}}

\defcitealias{Pascale2024a}{P24}
\defcitealias{Read2019}{R19}
\defcitealias{Hayashi2020}{H20}

\begin{document}

\title{Leo I: the classical dwarf spheroidal galaxy with the highest dark-matter density}
\titlerunning{}

\author{
R. Pascale\inst{1}, \thanks{\email{raffaele.pascale@inaf.it}}
C. Nipoti \inst{2}, 
F. Calura \inst{1}, \and
A. Della Croce \inst{1,2}}
\institute{
INAF - Osservatorio di Astrofisica e Scienza dello Spazio di Bologna, via Gobetti 93/3, 40129 Bologna, Italy 
\and
Dipartimento di Fisica e Astronomia 'Augusto Righi', Alma Mater Studiorum - Università di Bologna, via Gobetti 93/2, 40129 Bologna, Italy
}
\authorrunning{Pascale et al.}
\date{Received ...; accepted ...}
 
\abstract
{Dwarf spheroidal galaxies (dSphs) are known for being strongly dominated by dark matter (DM), which makes them convenient targets for investigating the DM nature and distribution. Recently, renewed interest in the dSph Leo I has resulted from claims suggesting the presence of a central supermassive black hole, with mass estimates that challenge the typical expectations for dSphs, which are generally thought to host intermediate-mass black holes (IMBHs). However, \cite{Pascale2024a} presented new upper limits on the black hole mass, which are consistent with the expected range for IMBHs, solving the concerns raised in previous studies. Building on the analysis of \cite{Pascale2024a}, we examine the DM properties of Leo I inferred from the state-of-the-art dynamical models of that paper. Our results indicate that Leo I is the galaxy with the highest DM density among the classical dSphs, with a central DM density (measured at a distance of $150\pc$ from the galaxy centre) $\rhoCL=35.5_{-4.7}^{+3.8}\times10^7\Msun\kpc^{-3}$. According to our model, the DM density profile has logarithmic slope $\gammaCL=-0.89_{-0.17}^{+0.21}$ at $150\pc$, in line with literature values. At smaller distances the DM distribution flattens into a core of constant density, with a core radius of $\rc=72^{+40}_{-32}\pc$. Combined with the small pericentric distance of Leo I's orbit in the Milky Way, the new estimate of $\rhoCL$ makes Leo I decisive in the study of the anticorrelation between pericentre and central DM density, and suggests that the anticorrelation could be significantly steeper and more pronounced than previously estimated. Finally, despite its DM dominance, Leo I does not emerge as the most favorable target for indirect DM detection: the inferred DM decay $D$ and annihilation $J$ factors, $\log D(0.5^{\circ}) [\GeV\cm^{-2}] = 17.94_{-0.25}^{+0.17}$ and $\log J(0.5^{\circ}) [\GeV^2\cm^{-5}]= 18.13_{-0.18}^{+0.17}$ are consistent with previous estimates and lower than the highest values measured in dSphs.}

\keywords{galaxies: individual: Leo I -  galaxies: dwarf - stars: kinematics and dynamics - - Cosmology: dark matter 
}
\maketitle

\section{Introduction}
\label{sec:intro}


According to the current $\Lambda$CDM cosmological model, dwarf galaxies play a crucial role in cosmic structure formation, with hierarchical evolution driven by the merging of smaller systems \citep{Lacey1993,Lacey1994}, and dwarf galaxies dominating the galaxy population \citep{Marzke1998}. As such, dwarf galaxies serve as key probes for cosmology, especially given also the increasing tensions observed at these smaller scales \citep{Bullock2017,Sales2022}. Among dwarfs, dwarf spheroidal galaxies (dSphs) satellites of the Milky Way (MW) are of particular importance due to their proximity to us and their significant dark matter (DM) content \citep{BattagliaNipoti2022}.  These two aspects combined make dSphs important testbeds for cosmological models and excellent sites for studying DM \citep{Gilmore2007}. As an example, due to the lack of emission in other bands apart from optical, dSphs are ideal targets to look for gamma-ray emission related to DM particles annihilation or decay \citep{Bonnivard2015,GeringerSameth2015,Evans2016,Boddy2020,Li2021,Andrade2024}.

Another example of the importance of dSphs in the study of DM is the core-cusp problem \citep{deBlok2009,Bullock2017}. While DM-only simulations predict that halos should develop cuspy central density profiles, observational studies of low surface brightness and gas-rich dwarf galaxies suggest that their DM distributions are shallower, with central cores of constant density. In dSphs, determining the inner DM density is more challenging due to their lack of gas, which leaves only stellar kinematics as tracers of the gravitational potential. As a result, different methods and assumptions in modeling stellar orbits can lead to contrasting conclusions, contributing to the ongoing tension on this scale with some analyses suggesting the presence of cored profiles, others consistent with cuspy distributions (\citealt{Walker2009b}; \citealt{Strigari2008}; \citealt{AmoriscoEvans2011}; \citealt{BreddelsHelmi2013}; \citealt{Strigari2017}; \citealt{Pascale2018}; \citealt{Read2019}, hereafter R19; \citealt{Hayashi2020}, hereafter H20; \citealt{Arroyo2025}). The origin of such cores remains debated, as it may depend on interaction between baryions and DM \citep{Read2005,Governato2012,NipotiBinney2015} or indicate a need for alternative DM models \citep{Governato2015,Hui2017}


Leo I is one of the most luminous and massive dSphs within the MW halo, with a stellar mass of $5.5\times10^6\Msun$ \citep{McConnachie2012}. 
It has a regular morphology, with an elliptical structure on the plane of the sky. The ellipticity is $e\equiv1-\frac{b}{a} = 0.31\pm0.01$ \citep{Munoz2018}, with $a$ and $b$ the semi-major and -minor axes, respectively. This value is comparable to that of other well-known dSphs. The heliocentric distance is $256.7\pm13.3\kpc$ \citep{Mendez2002, Bellazzini2004, Held2010, Pacucci2023}, which makes Leo I the most distant dSph currently identified as satellite of the MW. We will adopt $D=256.7\kpc$ throughout this analysis. 

The orbit of Leo I is well known, with \citet{Mateo2008} providing a robust estimate of its line-of-sight (l.o.s.) velocity ($\vlos\simeq282\kms$). \cite{Sohn2013} determined the bulk velocity in proper motion (PM) for the first time, using HST images over a 5-year baseline. The tangential velocity component of the galaxy is significantly large; however, it remains lower than its radial component, resulting in a fairly elliptical orbit. \cite{Sohn2013} showed that Leo I likely entered the MW about $2.33\pm0.21 \Gyr$ ago, having its first pericentric passage approximately $1.05\pm0.09\Gyr$ ago. More recently, \cite{Battaglia2022}, using {\it Gaia} eDR3 data, provided a much more precise estimate of the bulk PM for Leo I, allowing for a more accurate orbital calculation. Their results indicate a very small bulk PM, leading to a more radial orbit with a pericentric passage of $\rperi=35_{-20}^{+24}.\kpc$. This value represents one of the smallest pericentric distances observed among the known dSphs. 

For the classical dSphs orbiting the MW an empirical anticorrelation is found between their orbital pericentric radius and their central DM density $\rhoCL$,  conventionally measured at $150$ pc from the center of the dwarf \citep{Kaplinghat2019,Cardona2023,Andrade2024}. The physical interpretation of the existence of such anticorrelation is still debated, and it could either find its explanation within CDM models or alternative models of DM, such as self-interacting DM \citep{Correa2015}. In any case, given its small pericentric radius and being also one of the dSphs with the highest DM central density \citep[e.g.][]{Hayashi2020,Andrade2024}, the galaxy is an especially interesting object to study in this context.


Distant galaxies serve as valuable tools for investigating the distribution of DM within the outer halo of the MW, providing essential constraints on the potential of the MW at large distances. However, their analysis requires a cautious approach. For example, the discrepancy between the PM estimates of \cite{Sohn2013} and \cite{Battaglia2022} may lead to different dynamical mass estimates in the outer parts of the MW \citep{Battaglia2022}. Nonetheless, this impact should remain moderate since the overall velocity of Leo I is primarily influenced by its l.o.s. component. In any case, \cite{Battaglia2022} and \cite{Sohn2013}, along with estimates from \cite{Fritz2018} and \cite{Gaia2018b}, agree on the timing of the (only) pericentric passage, $\simeq1\Gyr$ ago.

Leo I underwent an extended star-forming phase, marked by recurring bursts lasting until approximately $1\Gyr$ ago \citep{RuizLara2021}, making it among the youngest MW dSphs \citep{Bellazzini2004}. Notably, the latest star formation episode ($1\Gyr$ ago) coincides with its closest approach to the MW centre, which supports the hypothesis that ram pressure stripping close to the MW center quenched the star formation, capturing Leo I in the transition from dIrr to dSph \citep{RuizLara2021}. 


Recently, Leo I has garnered renewed interest due to the possible detection of a supermassive black hole (SMBH) at its center, as reported by \cite{Bustamante2021}. This SMBH, with a mass intriguingly comparable to the total stellar mass of the system, would challenge established models of black hole (BH) formation and evolution. To reconcile this anomaly, \cite{Pacucci2023} proposed that Leo I was initially more massive, losing a significant portion of its mass due to tidal interactions with the MW, rendering the SMBH consistent with a more massive original system. However, this model is unconvincing due to the unrealistic orbit required and its inconsistency with the mass-metallicity relation. Recently, \citet[][hereafter P24]{Pascale2024a} reanalyzed Leo I, using the same dataset as \cite{Bustamante2021} but employing more sophisticated models and a statistically robust approach. This reanalysis excludes the presence of a SMBH and shows that Leo I’s central region could, at most, host an intermediate-mass black hole (IMBH), with an estimated upper limit on the mass $\log\MBH[\Msun]<5.27$ at $1\sigma$, thus suggesting that Leo I’s internal kinematics is actually consistent with a no-BH scenario.




In this work, we build upon the models presented by \citetalias{Pascale2024a}, shifting our focus toward a detailed analysis of the DM properties in Leo I, as the models presented by \citetalias{Pascale2024a} also provide accurate predictions for the DM distribution within the galaxy. Specifically, we investigate the inner structure of Leo I’s DM halo, aiming to constrain its inner slope. Furthermore, we reassess $\rhoCL$, the DM density within $150\pc$, and incorporate this revised estimate into the context of established scaling relations. We provide new estimates of the decay and annihilation $D$ and $J$ factors, essential quantities in the search for gamma-ray emission from DM particle interactions. 

This paper is organized as follows. In Section~\ref{sec:datamodels} we give a brief recap of the dataset and the models used by \citetalias{Pascale2024a}. In Section~\ref{sec:res} we discuss our results and compare them with the literature. Section~\ref{sec:concl} we draw our conclusions.

\begin{table}
\centering
\caption{Leo I main structural parameters.}
\renewcommand{\arraystretch}{1.3}
\begin{tabular}{ccc}
\toprule
Parameter & Value & Reference \\
\bottomrule
Ra & 10h 08m 28.1s & (a) \\
Dec & +12d 18m 23s & (a) \\
$D$ [kpc] & $256.7\pm13.3\kpc$ & (b) \\
$\MV$ & $-12.0\pm0.3$ & (a) \\
$\Mst$ [$\Msun$] & $5.5\times10^6$ & (a) \\
$e$ & $0.31\pm0.01$  & (c) \\
$\rc$ [$\amin$]& $3.3\pm0.3$ & (a) \\
$\rt$ [$\amin$] & $12.6\pm1.5$ & (a) \\
$\Reff$ [$\amin$] &  $3.65\pm0.03$ & (c) \\
\bottomrule
\end{tabular}
\tablefoot{From top to bottom: Right ascension and Declination (Ra, Dec); Distance $D$; V-Band absolute magnitude $\MV$; stellar mass $\Mst$; ellipticity $e$; core radius $\rc$ and truncation radius $\rt$ from one-component King models; effective radius $\Reff$. References: (a) \cite{McConnachie2012}; (b): \cite{Pacucci2023}; (c): \cite{Munoz2018}; (d): \cite{IrwinHatzidimitriou1995}}
\label{tab:parameters}
\end{table}

\section{Data and models}
\label{sec:datamodels}

In this section, we provide a brief summary of the dataset and the dynamical models used by \citetalias{Pascale2024a}. For an extensive discussion, we refer the reader to \citetalias{Pascale2024a}, and to their appendix A for a comprehensive description of the Bayesian strategy and likelihood methodology adopted, which we do not report here for the sake of conciseness.

\begin{table*}[h]
\centering
\caption{Inference on the models free parameters.}
\renewcommand{\arraystretch}{1.3}
\begin{tabular}{lccccc}
\toprule
Parameter & Prior & Median & $1\sigma$ & $3\sigma$ \\
\bottomrule
$\log\Mst$ [$\Msun$]        & [6.439, 7.041]      & 6.759 & [6.567, 6.939] & [6.441, 7.040] \\
$\log\Jcst$ [$\kpc\kms$]    & [-2, 0]  & -0.378 & [-0.453, -0.316] & [-0.698, -0.190] \\
$\log\Jst$ [$\kpc\kms$]     & [0, 1]   & 0.460 & [0.441, 0.479] & [0.401, 0.522] \\
$\log\Jtst^\ast$ [$\kpc\kms$]     & -  & - & - & - \\
$\hzst$                     & [0, 1.5] & 0.783 & [0.737, 0.825] & [0.639, 0.929] \\
$\gzst$                     & [0, 1.5] & 1.024 & [1.008, 1.040] & [0.977, 1.069] \\
$\Gammast$                  & [0, 3]   & 1.378 & [1.289, 1.461] & [1.040, 1.616] \\
$\Bst$                      & [3, 9]   & 6.576 & [6.336, 6.805] & [5.918, 7.295] \\
$\etast$                    & [0.5, 10]& 7.377 & [5.489, 8.979] & [3.315, 9.985] \\
$\alphast$ (fixed)   & -  & 0 & - & - \\
\midrule
$\log\Mdm$ [$\Msun$]        & [6, 11]  & 8.846 & [8.561, 9.053] & [8.091, 9.422] \\
$\log\Jcdm$ [$\kpc\kms$]    & [-1.5, 0]& -0.394 & [-1.089, 0.320] & [-1.497, 1.128] \\
$\log\Jdm$ [$\kpc\kms$]     & [0, 2.5] & 1.707 & [1.321, 2.170] & [0.335, 2.494] \\
$\log\Jtdm$ [$\kpc\kms$] (fixed)     & - & 1.92 &  - & -  \\
$\hzdm=\gzdm$               & [0, 1.5] & 0.927 & [0.557, 1.227] & [0.130, 1.416] \\
$\Gammadm$                  & [0, 3]   & 1.301 & [0.974, 1.567] & [0.146, 2.156] \\
$\Bdm$                      & [3, 10]  & 5.479 & [3.854, 7.803] & [3.011, 9.983] \\
$\etadm$                    & [0.5, 5] & 2.985 & [1.493, 4.310] & [0.522, 4.996] \\
$\alphadm$ (fixed)    & - & 1.93 &  - & -  \\
\midrule
$\log\MBH$ [$\Msun$]        & [1, 7]  & 3.894 & [2.105, 5.274] & [1.005, 5.835] \\
$c$                         & [25, 35] & 29.773 & [29.581, 29.952] & [29.449, 30.063] \\
\bottomrule
\end{tabular}
\tablefoot{List of free parameters alongside the adopted prior distributions (which are uniform in the reported interval), and the median values, $1\sigma$ and $3\sigma$ confidence intervals of the posterior distributions resulting from the analysis of \citetalias{Pascale2024a}. \\
${}^\ast$ Since $\alphast=0$. the value of $\Jtst$ is unimportant, see equation ~(\ref{for:df}).}\label{tab:fitparams}
\end{table*}

\subsection{Data}
\label{subsec:data}

The dataset integrates photometric and kinematic measurements. The photometric component is based on the surface brightness profile from \cite{Bustamante2021}, derived from SDSS g-band imaging. This profile extends from the central $30\pc$ to approximately $1.5\kpc$. Unlike the profiles adopted in previous studies, the surface brightness profile derived in this work both probes the innermost regions of Leo I and extends to significantly large projected radii. Also, in the central galaxy region, crowding and completeness was assessed using high-resolution HST images. However, it must be stressed that, despite the applied corrections, the use of wide-field, ground-based imaging for a galaxy at such a large heliocentric distance makes the computation of the surface brightness profile prone to potential biases and limitations. In particular, ground-based imaging of distant, low-surface-brightness dwarf galaxies is notoriously challenging due to the extremely low surface brightness of these systems, which often approaches the limiting sky background level.

For the inner galaxy regions, the models fit the central line-of-sight velocity distributions (LOSVDs) obtained from integral field spectroscopy data. These distributions provide dynamical information extending out to approximately $110\pc$ from the galaxy’s center. Finally, beyond $110\pc$, the authors used the galaxy's velocity dispersion profile computed from the sample of radial velocities of Leo I's members from \cite{Mateo2008}.


\subsection{Models}
\label{subsec:mod}
Leo I's models from \citetalias{Pascale2024a} are based on analytic distribution functions (DFs) depending on the action integrals $\JJ$ \citep{Binney2014,ColeBinney2017}. In their analysis, Leo I is represented as a multi-component galaxy, with DF-based stellar and DM components, and an optional central BH of mass $\MBH$ described by a Keplerian, fixed contribution to the total potential
\begin{equation}\label{for:phitot}
    \Phitot(r) = \Phist(r) + \Phidm(r) + \PhiBH(r),
\end{equation}
where $\Phist$, $\Phidm$ and $\PhiBH$ are the stellar, DM and BH potentials. For stars and DM the authors solve for the Poisson equation
\begin{equation}\label{for:poisson}
   \nabla^2\Phii(\bbx) = 4\pi G\int \d\bbv^3\fJi,
\end{equation}
where $i$ can be either stars ($i=\star$) or DM ($i={\rm DM}$). The DF of each component $\fJi$ is given by
\begin{equation}\label{for:df}
\begin{split}
 \fJi = & \fni\frac{\Mi}{(2\pi\Ji)^3} \biggl[1 - \beta\frac{\Jci}{\hJi} + \biggl(\frac{\Jci}{\hJi}\biggr)^2\biggr]^{-\frac{\Gammai}{2}} \times \\
 & \biggl[1 + \biggl(\frac{\Ji}{\hJi}\biggr)^{\etai}\biggr]^{\frac{\Gammai}{\etai}} \biggl[1 + \biggl(\frac{\gJi}{\Ji}\biggr)^{\etai}\biggr]^{\frac{\Gammai-\Bi}{\etai}} \times \\
 & e^{-\bigl[\frac{\gJi}{\Jti}\bigr]^{\alphai}},
\end{split}
\end{equation} 
where 
\begin{equation}\label{for:hg}
\hJi = \hri\Jr+\frac{3-\hri}{2}(|\Jphi|+\Jz)=\hri\Jr+\frac{3-\hri}{2}L   
\end{equation} 
and
\begin{equation}\label{for:hg2}
    \gJi = \gri\Jr+\frac{3-\gri}{2}(|\Jphi|+\Jz)=\gri\Jr+\frac{3-\gri}{2}L.
\end{equation} 
This DF framework allows for a variety of density profiles, including cored or cuspy density profiles, with different inner and outer velocity distributions. $\Mi$ is the total mass of the target component, $\Jci$, $\Ji$ and $\Jti$ are action scales that set the transition between different regimes in the action space (details are provided in \citealt{Vasiliev2019} and \citetalias{Pascale2024a}). $\Gammai$, $\Bi$, and $\etai$ are dimensionless parameters related to the inner and outer slopes of the density profile and the sharpness of the transition between the two.
The dimensionless parameters $\gri$ and $\hri$, with allowed ranges
$0<\gri<3$ and $0<\hri<3$, are  related to the model's velocity distributions. The models are spherically symmetric since the DFs~(\ref{for:df}) depend on the vertical ($\Jz$) and azimuthal ($\Jphi$) actions only through the angular momentum modulus $L=\Jz+|\Jphi|$.

In Table~\ref{tab:fitparams} we provide a convenient summary of the inference on the model's free parameters from \citetalias{Pascale2024a}, alongside the lower and upper bounds of the uniform prior distributions adopted in their Bayesian analysis. Here, it is sufficient to recall that confidence intervals on model parameters and any derived quantity are computed as percentiles from the corresponding marginalized distributions. Specifically, the median model corresponds to the 50th percentile, the $1\sigma$ confidence intervals are computed using the 16th and 84th percentiles, while the $3\sigma$ confidence intervals as the 0.15th and 99.85th percentiles.

\begin{figure*}
    \centering
    \includegraphics[width=1\hsize]{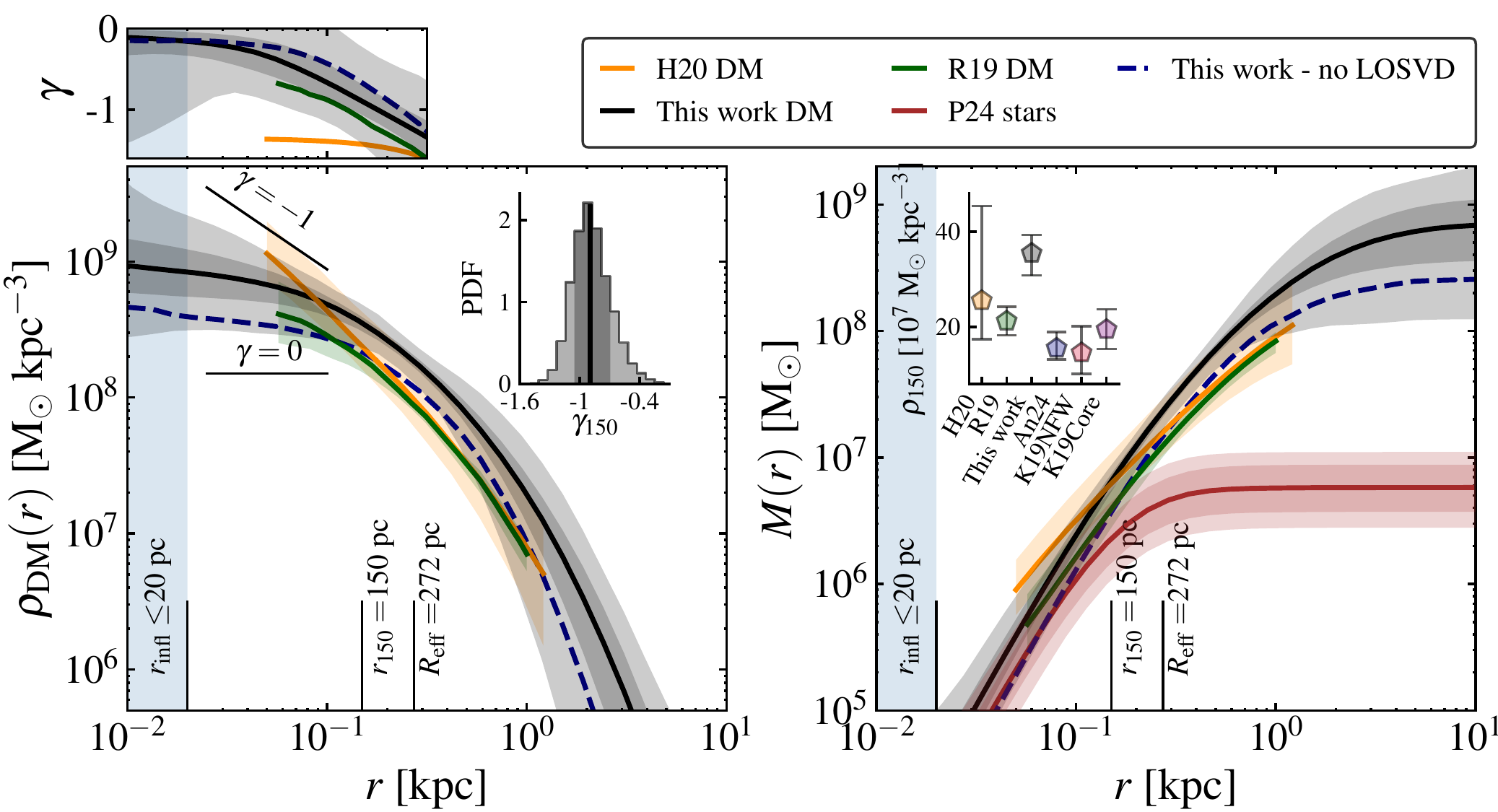}
    \caption{DM Properties. Left panel: DM density profile from the reference model (black solid line) with $1\sigma$ and $3\sigma$ confidence intervals (grey bands). The green and orange lines, together with the colored bands, show DM density profiles from \citetalias{Read2019} and \citetalias{Hayashi2020}, respectively, alongside their $1\sigma$ confidence intervals. The blue-dashed line is the median model obtained fitting only the l.o.s. velocity dispersion profile in Section~\ref{subsec:data}, with models~\ref{subsec:mod}. The black vertical line indicates the upper limit on Leo I's BH influence radius \citepalias[$\rinfl\le20\pc$;][]{Pascale2024a}, while other vertical lines mark $150\pc$ and the effective radius $\Reff=272\pc$ \citep{McConnachie2012}. The small panel to the top shows the corresponding logarithmic slope $\gamma\equiv{\dd\ln \rhodm}/{\dd \ln r}$. Right panel: same as the left panel, but showing the enclosed DM mass profile. The red line indicates the median stellar mass with $1\sigma$ and $3\sigma$ uncertainties with a red band. The inset in the left panel shows the posterior distribution on $\gammaCL$. The inset in the right panel shows a compilation of $\rhoCL$ (with $1\sigma$ errorbars) values from various studies: \citetalias{Read2019} and \citetalias{Hayashi2020} (color-coded as in main panels); \citet[][An24]{Andrade2024} in blue; and \citet[][Kap19]{Kaplinghat2019} for NFW and cored DM models in red and purple, respectively. Note that \citet{Kaplinghat2019} and \citet{Andrade2024} provide the values of $\rhoCL$, but not density profiles.}
    \label{fig:fig1_dmstdens}
\end{figure*}

\section{Results}
\label{sec:res}

In this section, we present and discuss our results, which are based on further analysis of the models of \citetalias{Pascale2024a}. \citetalias{Pascale2024a} only focused on the central BH properties, but the same models also account for the presence of DM (see Section~\ref{subsec:mod}), to which we shift our attention here.

\subsection{Dark matter density distribution}
\label{subsec:dmdens}

In Fig.~\ref{fig:fig1_dmstdens} we present an overview on the DM properties of Leo I based on our analysis. The left panel shows the median DM density profile computed from the posterior distribution presented by \citetalias{Pascale2024a} with the $1\sigma$ and $3\sigma$ confidence intervals, and, for comparison, the $1\sigma$ profiles estimated by \citetalias{Read2019} and \citetalias{Hayashi2020}. A vertical solid line to the left indicates the upper limit of the BH influence radius\footnote{The BH influence radius is estimated as the distance from the galaxy center where the l.o.s. velocity dispersion is equal to the BH circular velocity \citep{BinneyTremaine2008}.} estimated by \citetalias{Pascale2024a}, while additional vertical solid lines marking the radius of $150\pc$ - used for estimates of $\rhoCL$ - and the effective radius $\Reff=272\pc$ from \cite[see also Table~\ref{tab:parameters}]{McConnachie2012}. The small panel to the top shows the corresponding logarithmic slope $\gamma\equiv\frac{\dd\log\rhodm}{\dd\log r}$ profiles. The right panel displays the galaxy's enclosed DM mass distribution, and, for reference, we also include the stellar mass profile resulting from our fit. According to our analysis, Leo I is confirmed to be a DM-dominated galaxy, with a dynamical-to-stellar mass ratio ($\Mdyn/\Mst$) of $\simeq6.4$ within $\Reff$, and $\simeq32.5$ within the stellar truncation radius $\rt=940\pc$. These ratios correspond to dynamical masses of approximately $2.82_{-0.13}^{+0.13}\times10^7\Msun$ within $\Reff$ and $18.70_{-3.38}^{+3.79}\times10^7\Msun$ within $\rt$ (see Table~\ref{tab:addparams}).

\begin{figure*}
    \centering
    \includegraphics[width=1\hsize]{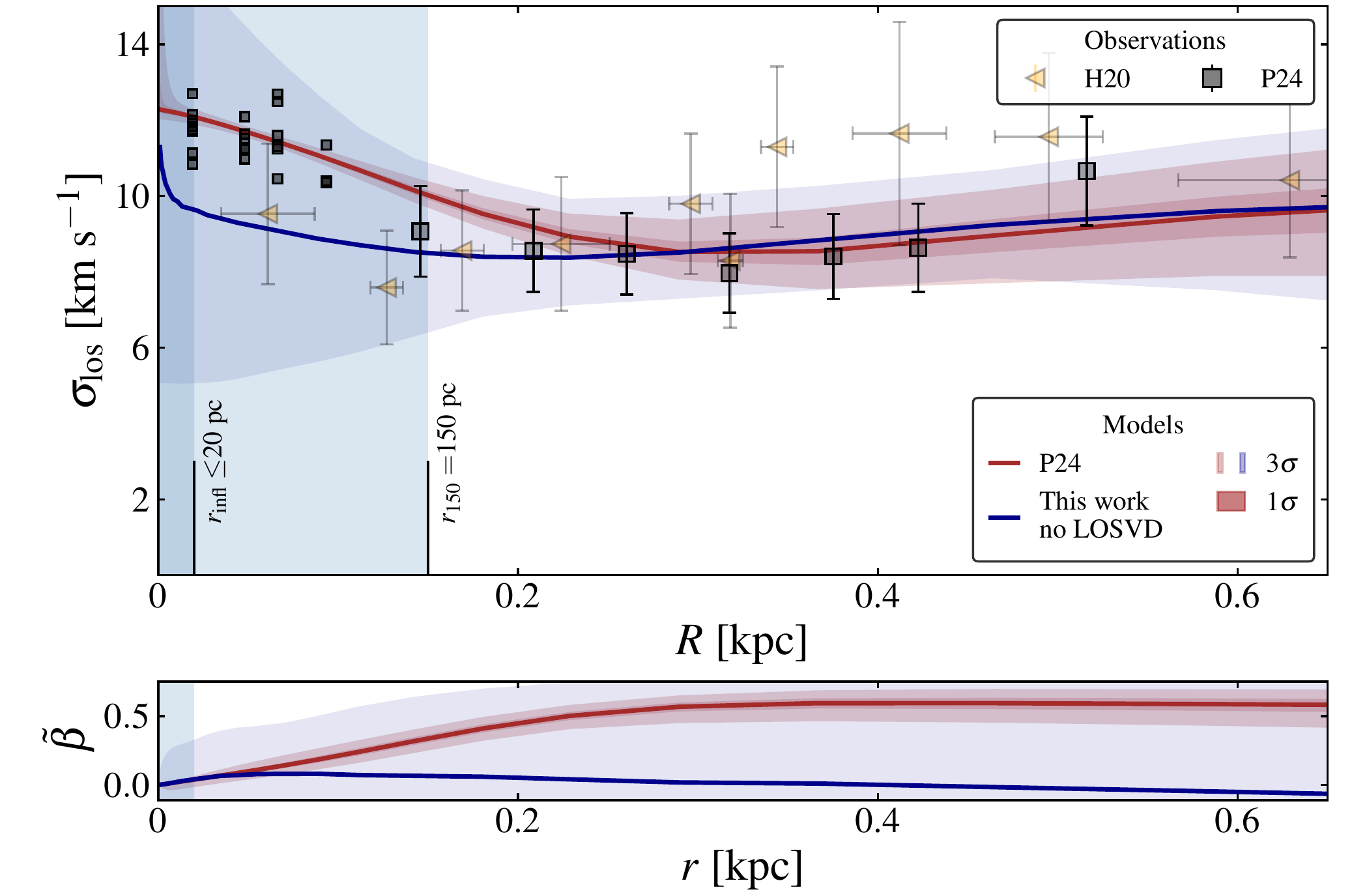}
    \caption{Kinematic properties of Leo I. Top panel: Median l.o.s. stellar velocity dispersion profile from \citetalias{Pascale2024a} (solid red line), with shaded regions representing the $1\sigma$ and $3\sigma$ confidence intervals. The blue solid line and the light blue band show, instead, the median l.o.s. velocity dispersion and the $3\sigma$ band obtained fitting the same dataset as \citetalias{Pascale2024a} except for the central LOSVDs, respectively. Black squares with error bars depict the l.o.s. velocity dispersion from \citetalias{Pascale2024a} computed with \cite{Mateo2008} data, while the smaller black squares in the inner region show the dispersion of the LOSVDs fitted by \citetalias{Pascale2024a} (see their figure 2). For comparison, l.o.s. velocity dispersion data from \citetalias{Hayashi2020} along Leo I’s major axis are shown as yellow triangles with errorbars. The light blue vertical bands mark the regions within Leo I's BH maximum influence radius ($\rinfl\le20\pc$) and within $150\pc$. Bottom panel: Symmetrized stellar anisotropy parameter profile derived from the median model of \citetalias{Pascale2024a}, including $1\sigma$ and $3\sigma$ confidence intervals. As in the top panel, the red curve and band correspond to the models from \citetalias{Pascale2024a}, while blue curve and band to the models obtained fitting the same dataset as \citetalias{Pascale2024a} except for the LOSVDs.}\label{fig:fig2_slos}
\end{figure*}

\citetalias{Hayashi2020} analyze the DM density distribution of all the classical dSphs using axisymmetric Jeans models. In their approach, both stars and DM are flattened, and they fit the radial distribution of stars alongside the l.o.s. velocity dispersion of stars along the major, minor, and an intermediate axes. The DM halo is represented with a generalized axisymmetric Hernquist profile \citep{Hernquist1990,Zhao1996}. This model allows for the description of halos with either cuspy central profiles or cored profiles, providing a smooth transition between these configurations. In their specific analysis, \citetalias{Hayashi2020} find a general preference for cuspy profiles in most of the dSphs studied, with Leo I being the one with the steepest inner slope (although with considerable uncertainties) with respect to better constrained cusps such as the ones the authors find in Draco or Ursa Minor. \citetalias{Read2019} employ a non-parametric spherical Jeans approach via the GravSphere \citep{ReadSteger2017} code to examine inner DM densities across a diverse sample of dwarf galaxies, including both dSphs and dwarf irregulars with varied star formation histories. Their results also indicate a preference for a cuspy profile in Leo I. In addition to the aforementioned studies, in an earlier work, \citet{Koch2007} investigated the DM density profile of Leo I. However, this study did not perform a systematic exploration of the parameter space, and the quality of the available kinematic data prevented the authors from drawing firm conclusions regarding whether a cuspy or cored DM profile is preferred, with both being compatible with the data.

In our analysis, we find that the DM density profile of Leo I asymptotically approaches a median slope of $\gamma=0$ for small distances (top-left panel of Fig.~\ref{fig:fig1_dmstdens}), indicating the presence of a core. This result is in apparent contrast with the cuspy profiles proposed in previous studies. Particularly, when compared to the sample of \citetalias{Hayashi2020}, our results diverge significantly, as their model predicts a slope of $\gamma=-1.35^{+0.32}_{-0.61}$, even steeper than a pure NFW with $\gamma=-1$, indicating a strongly cusped profile. By comparison, our results are more aligned with those of \citetalias{Read2019}. Although \citetalias{Read2019} report a cuspy profile, their spatial coverage of Leo I is more limited than ours, and within the same radial range, our results are compatible with theirs. As a reference, \citetalias{Read2019} measure a slope of $\gammaCL = -1.15^{+0.33}_{-0.37}$ at $r = 150\pc$, which is consistent within $1\sigma$ with our result of $\gammaCL = -0.89^{+0.21}_{-0.17}$. To provide a clear representation of the $\gammaCL$ parameter, in the small inset of Fig.~\ref{fig:fig1_dmstdens} we show its posterior. It is worth noting that the core-like nature of our profile, with a slope approaching zero, appears only in the very central regions, at distances smaller than or comparable to the innermost data points used by \citetalias{Read2019}. Also, the top-left panel of Fig.~\ref{fig:fig1_dmstdens} reveals that on the scales of $\rinfl$, although the median model suggests a central slope $\gamma$ close to zero, the distribution of $\gamma$ exhibits a highly asymmetric tail extending toward $\gamma \simeq -1$, with even smaller values at 3$\sigma$. This feature is particularly intriguing, and it arises from the inclusion of a central BH in the models, whose impact on the DM distribution is degenerate with, since it also produces, for $r\to0$, a cusp.

We provide a quantitative measure of the core extension in the central regions of the DM halo, defining its core radius, $\rc$, as the distance from the galaxy center at which the logarithmic slope of the DM density profile is $\gamma = -0.5$. We find $\rc=72^{+40}_{-32}\pc$, smaller indeed than $150\pc$, further supporting the consistency of our results with the findings of \citetalias{Read2019}, and marginally larger than the BH maximum influence radius $20\pc$. Additionally, the DM slope at the radius of influence is $\gamma_{\rinfl}=-0.16^{+0.12}_{-0.17}$, reinforcing the evidence that only the very innermost regions tend to a cored structure. These quantities are also reported in Table~\ref{tab:addparams} for reference.

The DM density profiles inferred from our models of Leo I consistently indicates a higher normalization across almost all radii when compared to previous profiles in the literature. This discrepancy is especially pronounced at $150\pc$, a reference distance used to assess the amount of DM in the centre of dSphs, and thus a scale that is well-constrained by several studies. When compared to \citetalias{Hayashi2020}, who provide a comprehensive estimate of $\rhoCL$ across all classical dSphs, Leo I emerges as the most DM-dense classical dSph. Although \citetalias{Hayashi2020} also reported that Leo I has the highest $\rhoCL$, our analysis yields a value that is almost twice as large, but still consistent within $1\sigma$, reinforcing its status as the densest dSph. The small inset in the right-hand panel of Fig.~\ref{fig:fig1_dmstdens} presents a collection of $\rhoCL$ from various studies, including those by \citetalias{Hayashi2020} and \citetalias{Read2019}, along with additional estimates from \cite{Kaplinghat2019} and \cite{Andrade2024}, with \cite{Kaplinghat2019} providing two measurements based on different DM models, one assuming a core and the other a cusp. This comparison highlights that the values from \citetalias{Hayashi2020} has considerably higher uncertainty than all other studies and that our measurement indicates a DM density at $150\pc$ that is at least twice as high as any previous estimate.

A notable aspect of this discrepancy is that all studies, except for \citetalias{Pascale2024a}, agree with each other within the error bars. Interestingly, these works are consistent despite employing different methods. For instance, \citetalias{Hayashi2020} use flattened models and compute three separate profiles along the minor, major, and intermediate axes of the galaxy, fitting them with axisymmetric Jeans models. Both \citetalias{Read2019} and \cite{Kaplinghat2019} fit higher moments of the l.o.s. velocity distribution with spherical Jeans models, while \cite{Andrade2024} adopt DF-based models, as in our work.

The major distinction between the other studies and ours lies in the employed dataset. While all analyses rely on the l.o.s. velocity data from \cite{Mateo2008}, \citetalias{Pascale2024a} incorporates an additional dataset: the central LOSVDs data from \cite{Bustamante2021}. In Fig.~\ref{fig:fig2_slos}, we present the velocity dispersion profile derived by \citetalias{Pascale2024a} (black squares with error bars) alongside the median l.o.s. velocity dispersion model and confidence intervals from \citetalias{Pascale2024a}. For comparison, we also show the velocity dispersion profiles from \citetalias{Hayashi2020} as yellow points. The profiles from \citetalias{Hayashi2020} and \citetalias{Pascale2024a} align closely for $r > 150\pc$, while they diverge within $150\pc$, where \citetalias{Pascale2024a} incorporate the central LOSVDs into the fit. These LOSVDs exhibit velocity dispersions that reach up to $12\kms$ (shown as small black points in Fig.~\ref{fig:fig2_slos}), a value higher than the inner profile from \citetalias{Hayashi2020}, which stabilizes around $8\kms$. The bottom panel of Fig.~\ref{fig:fig2_slos} shows the symmetrized anisotropy parameter $\tbeta$ defined as \citep{Read2006}
\begin{equation}
    \tbeta = \frac{\sigmar^2-\sigmat^2}{\sigmar^2 + \sigmat^2},
\end{equation}
with $\sigmar$ and $\sigmat$ the radial and tangential velocity dispersions, respectively. The parameter $\tbeta$ it is related to the well known anisotropy parameter $\beta$ \citep{BinneyTremaine2008}
\begin{equation}
    \tbeta = \frac{\beta}{2-\beta}.
\end{equation}
The advantage of using $\tbeta$ over $\beta$ is that $\tbeta$ is symmetric between $-1$ and 1. $\tbeta$ measures tangential anisotropy when it lies between $-1$ and 0 (whereas $\beta<0$). Both $\tbeta$ and $\beta$ are equal to 0 for isotropy, and both $\tbeta$ and $\beta$ correspond to radial anisotropy in the range between 0 and 1.

We believe that the inclusion of the additional LOSVD dataset is the primary reason for our higher inferred DM density. To test this hypothesis, we repeated the identical Bayesian model-data comparison performed by \citetalias{Pascale2024a}, but excluding the central LOSVDs. The resulting median DM density, mass and logarithmic slope profiles are reported in  Fig.~\ref{fig:fig1_dmstdens}. As shown, the DM density profile is fully consistent with those derived in other studies, so does the value of $\rhoCL$. In this model, the dynamical mass within $\Reff$ and $\rt$ decreases by nearly a factor of two, becoming $1.94_{-0.38}^{+0.55}\Msun$ and $11.54_{-4.02}^{+7.48}\Msun$, respectively. These revised values are in agreement with all previous estimates, including that by \citet{Koch2007}. While \cite{Koch2007} provide an estimate of the DM mass within $\rt$, and we report the total dynamical mass, the system at this radius is effectively DM-dominated, making the two measurements directly comparable. 

This behavior can be explained qualitatively as follows, although the interplay between the models' free parameters is not necessarily straightforward. The central LOSVDs, whose dispersions are particularly high when compared to the outer l.o.s. velocity dispersion profile, increase the amount of mass required in these regions and. This leads to an overall increase of DM density for a fixed anisotropy. At the same time, to fit the outer l.o.s. velocity dispersion profile, a strong radial anisotropy in the outer regions is favored (see the bottom panel of Fig.~\ref{fig:fig2_slos}). This is because radial anisotropy results into an increase of the velocity dispersion along the radial component, which in turn decreases the l.o.s. velocity dispersion profile in outskirts \citep{Pascale2019}. 

Residual differences between our inferred DM density profiles and those reported in previous studies may be attributed to the use of slightly different surface brightness or stellar density profiles by various authors. As discussed in Section~\ref{subsec:data}, the surface brightness profiles commonly adopted in the modeling of dSphs are derived from ground-based observations, which inevitably carry systematic biases and limitations. Moreover, unlike other works where these profiles serve as fixed inputs to the models, in our approach they are fitted simultaneously with the kinematic data, properly accounting for observational uncertainties, which may also contribute to these discrepancies.

Regarding the large $1\sigma$ confidence levels in the mass and density (and, thus, $\rhoCL$) distributions reported by \citetalias{Hayashi2020}, we believe that these arise from their use of flattened models, which allow for marginalization over the flattening of the DM halo. This approach likely increases the uncertainty in the inferred density estimates, resulting in the broader error ranges measured in their study.

\begin{figure}
    \centering
    \includegraphics[width=1\hsize]{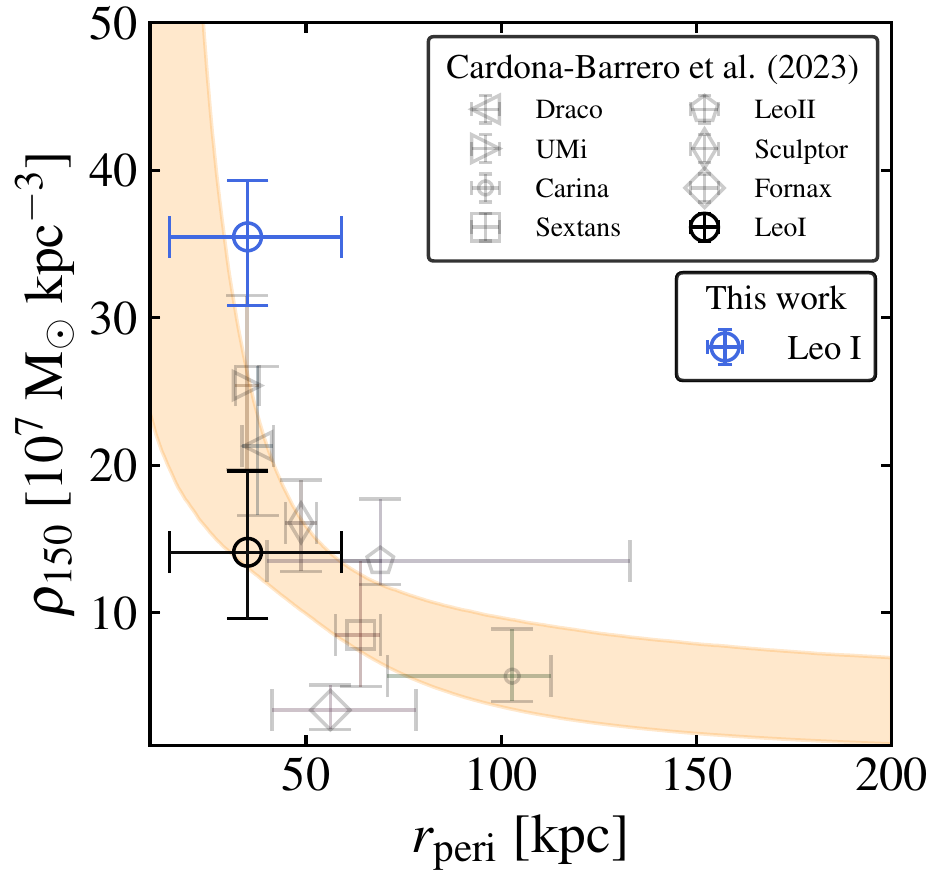}
    \caption{$\rperi$ - $\rhoCL$ anticorrelation for the eight classical dSphs from \citet{Cardona2023}. Central $\rhoCL$ densities are taken from \citet{Kaplinghat2019}, while pericentric distances are from \citet{Battaglia2022}. The orange band represents the fit to the grey and black points as derived by \citet{Cardona2023}. The blue point corresponds to the measurement for Leo I, with $\rhoCL$ derived in this work.}
    \label{fig:fig3_rperi}
\end{figure}


\begin{table}[h]
\centering
\caption{DM and stellar parameters computed from the posterior of \citetalias{Pascale2024a}.}
\renewcommand{\arraystretch}{1.5}
\begin{tabular}{lc}
\toprule
 Parameter & $1\sigma$ \\
\bottomrule
$\Mdyn/\Mst(\Reff)$ & $6.4_{-2.2}^{+3.6}$ \\
$\Mdyn/\Mst(\rt)$ & $32.5_{-11.0}^{+18.2}$ \\
$\Mdyn(\Reff)$ [$10^7\Msun$] & $2.82_{-0.13}^{+0.13}$ \\
$\Mdyn(\rt)$ [$10^7\Msun$] & $18.70_{-3.38}^{+3.79}$ \\
$\rhoCL$ [$10^7\Msun\kpc^{-3}$] & $35.5_{-4.7}^{+3.8}$  \\
$\rc$ [pc] & $72_{-32}^{+40}$ \\
$\gamma_{\rinfl}$ & $-0.16_{-0.17}^{+0.12}$ \\
$\gammaCL$ & $-0.89_{-0.17}^{+0.21}$ \\
$\log D(0.5^{\circ})$ [$\GeV\cm^{-2}$] & $17.94_{-0.25}^{+0.17}$  \\
$\log J(0.5^{\circ})$ [$\GeV^2\cm^{-5}$] & $18.13_{-0.18}^{+0.17}$ \\
\bottomrule
\end{tabular}
\tablefoot{From top to bottom: dynamical-to-stellar mass ratio within the stellar effective radius, $\Mdyn/\Mst(\Reff)$, and within the stellar truncation radius, $\Mdyn/\Mst(\rt)$; dynamical mass within the stellar effective radius, $\Mdyn(\Reff)$, and within the stellar truncation radius $\Mdyn(\rt)$; DM density at 150 pc, $\rhoCL$; core radius of DM halo, $\rc$, logslope $\gamma$ measured at the BH's maximum influence radius, $\gamma_{\rinfl}$, and at 150 pc, $\gammaCL$; DM decay $D$-factor within $0.5^{\circ}$, $\log D(0.5^{\circ})$, and DM annhilation $J$-factor within $0.5^{\circ}$, $\log J(0.5^{\circ})$.}\label{tab:addparams}
\end{table}

\subsection{Central density versus pericentric distance anticorrelation}
\label{subsec:rperi}

\begin{figure*}
    \centering
    \includegraphics[width=1\hsize]{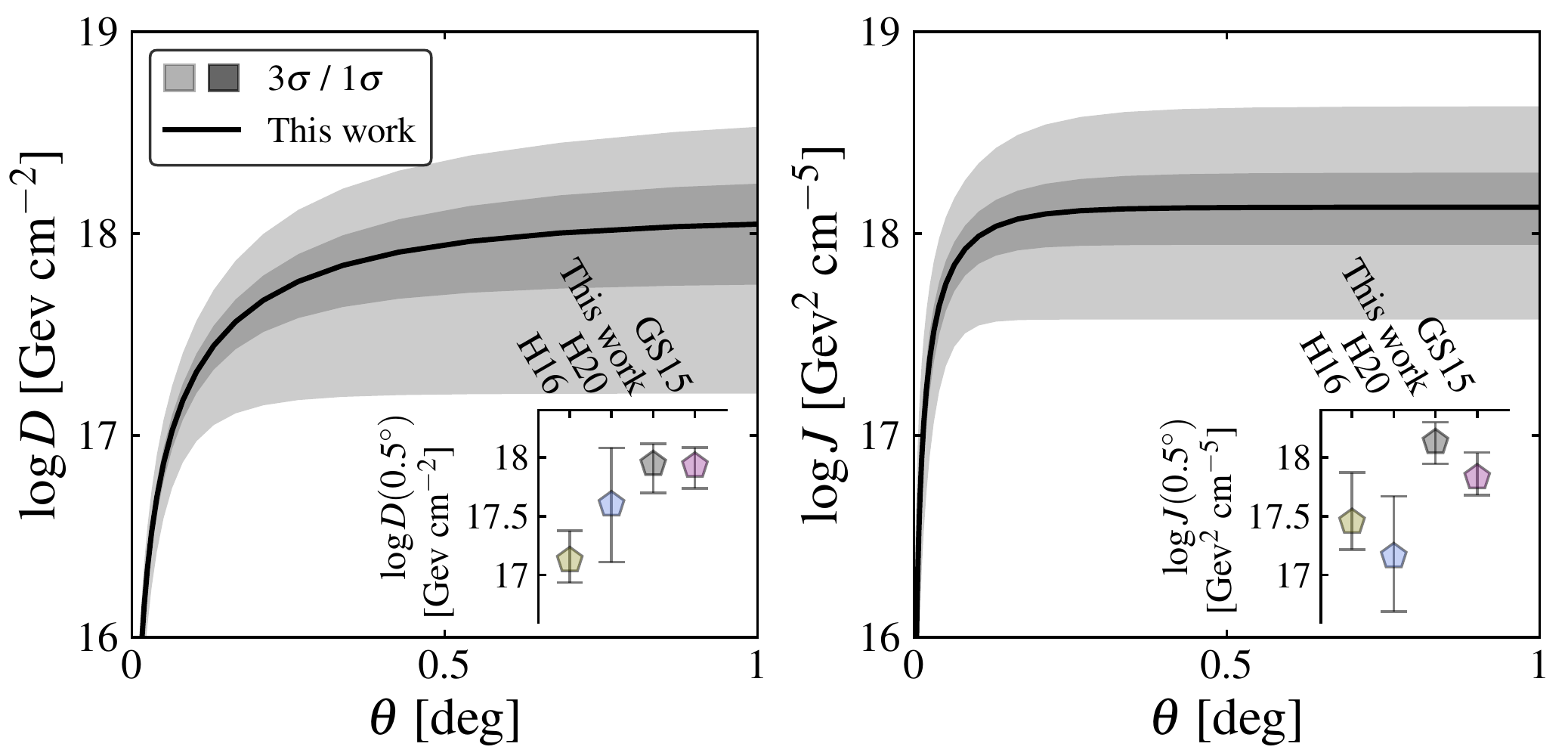}
    \caption{Left panel: DM decay $D$-factor computed from the reference model of \citetalias{Pascale2024a} (black solid line) together with the $1\sigma$ and $3\sigma$ bands. Right panel: same as the left panel but showing  the model's $J$-factor profile. The small insets in the bottom panel show the models $D$ and $J$ factors at an aperture of $0.5^{\circ}$ compared to values from the literature: \citet[][H16]{Hayashi2020} in yellow, \citet{Hayashi2020} and \citet[][GS15]{GeringerSameth2015} in purple.}
    \label{fig:fig4_jdfactors}
\end{figure*}

The advent of Gaia Data Release 2 (GDR2) has provided precise proper motion measurements of the bulk velocity of dSphs, significantly improving the reconstruction of their orbits within the Milky Way potential. This led several studies to identify a potential anticorrelation between the pericentric distance, $\rperi$, and the central DM density, $\rhoCL$, of the classical dSphs \citep{Kaplinghat2019,Cardona2023}. Such a trend implies that dSphs that pass closer to the Galactic center tend to have higher central DM density. However, the existence of this anticorrelation remains highly debated, and its underlying physical origin is still uncertain. A few mechanisms have been proposed to explain it.

For instance, the so-called "survivor bias" model (\citetalias{Hayashi2020}) posits that low-density dSphs with small pericentric distances have been disrupted by tidal interactions, leaving behind only the densest systems. However, this scenario fails to explain why galaxies with high $\rhoCL$ and large $\rperi$ are absent. Instead, some authors suggest that the anticorrelation could naturally arise in alternative DM scenarios, due to intrinsic and different properties of DM in non-standard scenarios. For instance, the anticorrelation might arise as a consequence of tidal-stripping-driven gravothermal core collapse of the satellite dwarf halos \citep{Correa2015} in the context of self-interacting DM (SIDM).  Others argue that the anticorrelation is merely an artifact of incompleteness or biased parameter estimates, rather than a genuine physical relationship. The anticorrelation is particularly controversial, as no similar trend has been observed in the population of ultrafaint dwarf galaxies, also satellites of the Milky Way.

In a recent work, \citet{Cardona2023} gathered different sets of central DM densities and pericentric distances from various studies \citep{Kaplinghat2019,Hayashi2020,Battaglia2022}, and explored multiple combinations of these datasets to assess the statistical occurrence of the proposed anticorrelation. Their analysis also incorporates $\rperi$ estimates that account for the time-dependent Milky Way potential, including the influence of the Large Magellanic Cloud’s infall, in order to marginalize and examine potential biases arising from the use of specific datasets and different assumptions. The results indicate that the anticorrelation is statistically significant in only 12\% of the analyzed dataset combinations, suggesting that evidence for this relationship is, apparently, weak and highly sensitive to the choice of datasets.

In Fig.~\ref{fig:fig3_rperi}, we show an illustrative example of the 24 $\rhoCL$–$\rperi$ combinations explored by \citet{Cardona2023}, along with their fitted anticorrelation (the "linear" fit proposed by the authors is in log-log space). While this is not the case where the anticorrelation is intrinsically the most significant, it remains detectable at the 1$\sigma$ level according to \citet{Cardona2023}. Our revised $\rhoCL$ estimate (blue point) further strengthens this trend, shifting the innermost data point upward. This is a particularly striking aspect, as it significantly enhances the anticorrelation, making it much more pronounced. This adjustment may increase the statistical significance of the anticorrelation in dataset combinations that previously showed weak or no evidence for it.

\subsection{$J$ and $D$ factors}
\label{subsec:jd}

The $J$- and $D$- factors are key quantities for indirect DM detection experiments. These experiments focus on searching for gamma-ray emission resulting from DM annihilation ($J$-factor) and decay ($D$-factor) within systems with possibly a large amount of DM \citep[e.g.][]{Zeldovich1980,Turner1984}. In this respect, dSphs are of particular interest since they are DM-dominated, but also they lack other sources in the gamma-ray band, and, additionally, they are relatively close to Earth. To date, no significant gamma-ray excess has been detected from dSphs, and all constraints on DM particle properties are based on non-detections. These non-detections have been used to place upper limits on the annihilation cross-sections and particle masses of DM candidates, with some studies ruling out certain classes of particles as viable explanations for the observed data \citep{Ackermann2015,GeringerSameth2015,Hoof2020,DiMauro2022,Crnogorcevic2024}. Despite the absence of direct signals, these galaxies remain prime targets for future DM searches.

The $J$ factor quantifies the probability of detecting a gamma-ray signal resulting from the annihilation of DM particles. Similarly, the $D$ factor relates to the probability of gamma-ray emission due to the decay of DM particles. These factors are defined as \citep{Evans2016}
\begin{equation}\label{for:j}
    J(\theta) = \frac{2\pi}{d^2}\int_{-\infty}^{+\infty}\dd z\int_0^{\theta d}\rhodm^2(R, z)R\dd R,
\end{equation}
and 
\begin{equation}\label{for:d}
    D(\theta) = \frac{2\pi}{d^2}\int_{-\infty}^{+\infty}\dd z\int_0^{\theta d}\rhodm(R, z)R\dd R,
\end{equation}
where $d$ is the distance of the galaxy, while $\theta$ is the angular distance measured from the center of the galaxy, $z$ is the l.o.s. while $R$ is the distance from the center of the galaxy on the plane of the sky. It is essential to note that the $J$ and $D$ factors are determined by two main factors: the DM density profile of the galaxy and its distance from the Earth. Leo I is particularly interesting as it is the most DM-dense dSphs, yet it is also the most distant, leading to some trade-offs in terms of its DM signature. 

Fig.~\ref{fig:fig4_jdfactors} shows the profiles of the $J$ and $D$ factors for Leo I as a function of the angular distance from the galaxy center, computed based on the posterior distributions from \citetalias{Pascale2024a}. Typically, $J$ and $D$ factors are calculated within a region of $0.5\deg$, which corresponds to the resolution of the Fermi LAT telescope. In Fig.~\ref{fig:fig4_jdfactors}, the small insets also provide comparisons with values from previous studies in the literature. While the new estimate of the $D$ is generally consistent within the error bars with previous estimates, the values of the $J$ factor differ significantly. Our new estimate of the $J$ factor is more than one sigma higher than the values reported by \cite{Hayashi2016,Hayashi2020} and is marginally compatible with \cite{GeringerSameth2015}. This discrepancy in the $J$ factor is likely due to the quadratic dependence on the DM density profile, which may accentuate differences in the $J$ factor for a given distance from the galaxy's center more than in the $D$ factor.

Compared to other dwarf galaxies, our new estimate of the $J$ factor places Leo I as a more favorable target than previously suggested. However, it is still not among the most promising dSphs for DM searches. Even though recent studies show that galaxies such as Fornax \citep{Pascale2019} and Sculptor \citep{Arroyo2025} exhibit systematically higher $J$ and $D$ factors than previously estimated (see also \citealt{Nipoti2024}), other galaxies, such as Draco and Ursa Minor, remain among the most favorable targets for DM detection \citep[see, for instance, ][]{GeringerSameth2015}.

\section{Conclusions}
\label{sec:concl}

This study provides a revised characterization of the DM properties of the dSph Leo I. Building on the results of \citetalias{Pascale2024a}, we have conducted an in-depth analysis of the DM distribution in the galaxy. The models here presented leverage state-of-the-art dynamical models based on $\fJ$ DFs and fit the central galaxy LOSVDs from \cite{Bustamante2021}, as well as photometric and outer spectroscopic measurements from \cite{Bustamante2021} and \cite{Mateo2008}. While \citetalias{Pascale2024a} primarily investigated the presence of a possible IMBH in Leo I, this study shifts the attention to the DM properties inferred from the same models. According to our new estimate, Leo I is the classical dSph known in the MW with the highest DM density. As a reference, we provide a central DM density of $\rhoCL=35.5_{-4.7}^{+3.8}\times10^7\Msun\kpc^{-3}$ at $150\pc$ from the galaxy center.

This new estimate has important implications for the anti-correlation between central DM density and the pericentric distance of MW satellites, whose physical origin remains an open question. For instance, DM-only simulations suggest a strong anti-correlation \citep{Hayashi2016}, whereas hydrodynamical simulations incorporating baryonic effects indicate that the MW’s disk potential can significantly alter satellite densities, leading to lower densities in satellites with smaller pericenters \citep{Robles2019}. While \cite{Cardona2023} recently demonstrated that the existence and strength of this anti-correlation depend on the used dataset - with a clear signal emerging only in certain cases - our revised and larger estimate of Leo I's DM density could strengthen the observed trend.

Our results on the central DM density of Leo I are also relevant to SIDM models. \citet{Correa21}, exploring models of SIDM with a velocity-dependent cross section, finds that Leo I favours a SIDM cross section larger than found for other dSphs with similar velocity dispersion. Using our estimate of $\rhoCL$ for Leo I, which is higher than that of \citet{Kaplinghat2019}, used by \citet{Correa21}, this conclusion is strengthened, because more efficient gravothermal collapse is required to produce a higher central DM density. \citet{Ebisu22} show that SIDM models with velocity-independent cross section are hard to reconcile with  the observed $\rhoCL$-$\rperi$ anticorrelation, because they predict dwarf galaxies with lower $\rhoCL$ than observed in classical dSphs, with Leo I being one of the galaxies with the strongest discrepancy: this discrepancy worsens if the central DM density of Leo I is as high as estimated in our work. 

We also explored the structure of the DM distribution in the context of the core-cusp problem. Our results indicate that Leo I's internal kinematics is compatible with the presence of a central DM core. We measure a core size (the distance from the galaxy center where the logslope $\gamma=-0.5$) of $\rc=72^{+40}_{-32}\pc$, larger than the maximum sphere of influence of a putative IMBH ($\rinfl=20\pc$). Here, we measure a logarithmic slope $\gamma_{\rinfl}=-0.16^{+0.12}_{-0.17}$. At larger radii, which correspond to the innermost regions covered by data in several studies from the literature, the DM density profile exhibits a mild cusp, with a logarithmic slope of $\gammaCL=-0.89_{-0.17}^{+0.21}$ at $150\pc$.

Finally, we provide new estimates of the $J$ and $D$ factors for Leo I, obtaining $\log D(0.5^{\circ}) [\GeV\cm^{-2}] = 17.94_{-0.25}^{+0.17}$ and $\log J(0.5^{\circ}) [\GeV^2\cm^{-5}]= 18.13_{-0.18}^{+0.17}$. Although these values are higher than some previous estimates, they remain within the range of literature values for Leo I. Thus, despite its status as a highly DM-dominated system, Leo I does not emerge as one of the most promising dSph targets for indirect DM detection via gamma-ray emission. 

A key difficulty in determining the surface brightness profiles of dSph galaxies lies in the use of wide-field ground-based imaging, where issues such as crowding, background subtraction, and the trade-off between field of view and spatial resolution, hence uniform coverage of the target, become significant. As a result, dynamical models relying on these profiles inherently carry a level of uncertainty associated with how the profiles themselves are constructed. Euclid will substantially mitigate these limitations: its imager combines a $0.57\deg$ field of view with spatial resolution (FWHM) $\simeq0.14\asec$ in VIS (I band) and $\simeq0.45\asec$ in NISP \citep[Y, J, H bands;][]{Cuillandre2025}, and a diffraction-limited PSF, enabling uniform, background-limited photometry down to I band magnitude $\simeq30$ mag $\asec^{-2}$ \citep{Hunt2025}. For Leo I, this translates into: (i) continuous coverage of the target beyond $40\asec$ in a single pointing; (ii) lower crowding and thus deeper stellar photometry; and (iii) self-consistent photometric calibration on a single detector, which minimizes completeness variations. Euclid is therefore expected to revolutionize the field of dynamical modeling by providing unprecedentedly high-quality data for Leo I, as well as for dSph galaxies and, generally, low-surface-brightness galaxies.

\begin{acknowledgements}
We are grateful to the referee, Prof. G. Gilmore, for the useful comments and feedback that helped improve the quality of the paper. We also thank Dr. Francesca Annibali for her constructive feedback and suggestions. This paper is supported by the Italian Research Center on High Performance Computing Big Data and Quantum Computing (ICSC), project funded by European Union - NextGenerationEU - and National Recovery and Resilience Plan (NRRP) - Mission 4 Component 2 within the activities of Spoke 3 (Astrophysics and Cosmos Observations). The research activities described in this paper have been co-funded by the European Union - NextGeneration EU within PRIN 2022 project n.20229YBSAN - Globular clusters in cosmological simulations and in lensed fields: from their birth to the present epoch.
\end{acknowledgements}


\bibliographystyle{aa}
\bibliography{main}

\end{document}